\documentstyle[11pt,newpasp,twoside,epsf]{article}

\begin{document}



\title{
The GMOS integral field unit: first integral field spectroscopy
with an 8m telescope}

\author{
Jeremy Allington-Smith\footnotemark[1],
Graham Murray, Robert Content, 
George Dodsworth, 
Roger Davies
}

\affil{Astronomical Instrumentation Group, University of Durham,\\
South Rd, Durham DH1 3LE, UK}

\author{Bryan W. Miller, James Turner}
\affil{Gemini Observatory, La Serena, Chile}

\author{Inger Jorgensen}
\affil{Gemini Observatory, Hilo, Hawaii, USA}

\author{Isobel Hook\footnotemark[2]}
\affil{UK Gemini Project Office, University of Oxford, Oxford, UK}

\author{David Crampton, Richard Murowinski}
\affil{Herzberg Institute of Astrophysics, Victoria, BC, Canada}

\footnotetext[1]{e-mail: j.r.allington-smith@durham.ac.uk}
\footnotetext[2]{address: Gemini North Observatory}

\begin{abstract}

The Gemini Multiobject Spectrograph (GMOS) installed on the Gemini-North
telescope has a facility for integral field spectroscopy over the wavelength
range $0.4-1.0\mu$m. GMOS is converted to this mode by the remote insertion
of an integral field unit (IFU) into the beam in place of the masks used for
multiobject spectroscopy.  With the IFU deployed, integral field
spectroscopy is available over a fully-filled contiguous field of 5 $\times$
7 arcsec with a sampling of 0.2 arcsec. A separate field of half the area,
but otherwise identical, is also provided to aid background subtraction. The
IFU contains 1500 lenslet-coupled fibres and is the largest-format
fibre-based IFU yet tested on the sky and the first facility of any type for
integral field spectroscopy employed on an 8/10m telescope.  We describe the
IFU and present results from commissioning.

\end{abstract}


\section{Introduction}
\label{intro}

The first of the two Gemini Multiobject Spectrographs has been completed and
tested on the Gemini-North Telescope. This paper describes the integral
field unit (IFU) which converts this multiobject spectrograph into an
integral field spectrograph.

The capabilities of GMOS (Fig. 1) can be summarised as follows.

\begin{itemize}
\item 0.07 arcsec/pixel image scale
\item $5.5 \times 5.5$ arcmin field
\item 0.4 - 1.1$\mu$m wavelength coverage
\item Resolving power up to $R = 10,000$ (with 0.25 arcsec slits)
\item Detector: CCD mosaic of $3 \times (4608 \times 2048$) pixels
\item Active control of flexure
\item On-instrument WFS for accurate guiding
\item Main modes:
\begin{itemize}
\item Multiobject using laser-cut slit masks
\item Integral field spectroscopy mode
\item Longslit spectroscopy using generic masks
\item Imaging 
\end{itemize}
\end{itemize}

\begin{figure}[tb]
\vspace{0mm}
\plotone{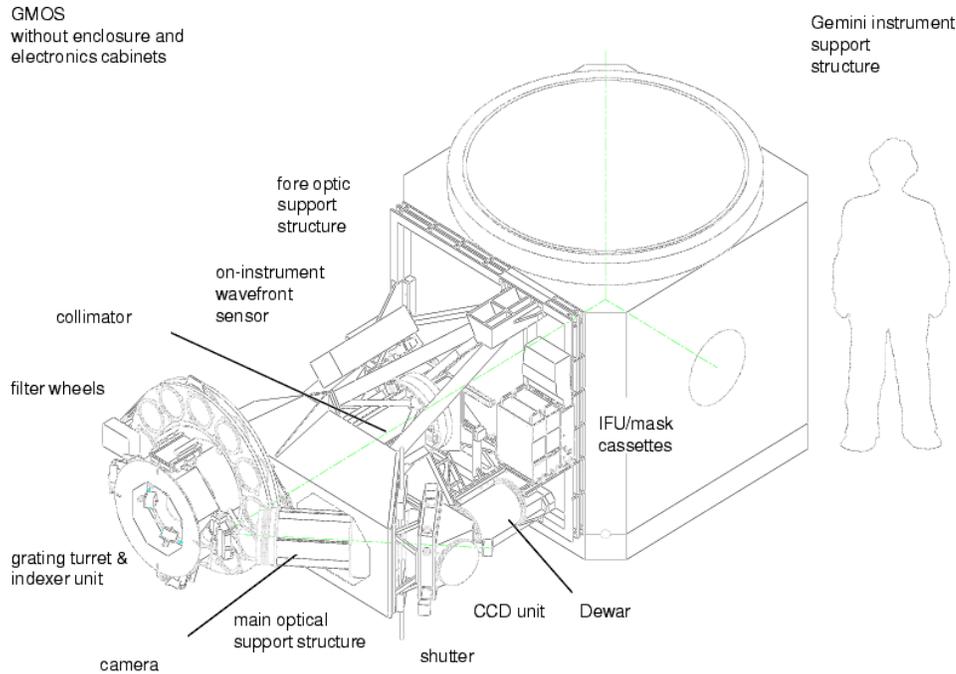}
\caption{Mechanical layout of GMOS without its enclosure and electronics
cabinets mounted on
the instrument support structure (ISS) of the Cassegrain focus. It is shown in
a sideways-looking mounting but may be mounted in an upward-looking
orientation by attaching it to the lower face of the ISS.}
\end{figure}

The scientific requirements were defined by the Gemini project. These are 
summarised below, where the adopted top-level specification and main
design features are also given.

\begin{itemize}

\item  To exploit good images from GEMINI, the IFU has a sampling
of 0.2 arcsec.
 
\item To provide unit filling factor in the field, the fibres are
coupled to a close-packed lenslet array at the input

\item To provide the largest possible contiguous field, 1000 fibers
were used to give an object field of $7 \times 5$~arcsec (1000 fibres)

\item To make provision for accurate background subtraction, 
an extra $5 \times 3.5$~arcsec field offset by 60~arcsec from the object
field is provided for background estimation (500 fibres).

\item To allow  a transparent change between slit and integral field modes,
the IFU is deployed by the mask exchanger and the input and output foci are
coplanar with the masks.

\item To obtain high throughput,  the fibres are lenslet-coupled at the
output and input to convert the F/16 beam to $\sim$F/5 for efficient use
with fibres.

\item The adoption of a low risk construction technique (to reduce risk to
schedule) dictated the use of the fibre+lenslet technique in preference to
image slicing (see Allington-Smith \& Content 1999 --- hereafter AC --- for
a summary of the techniques available)

\end{itemize}

\section{Description}

The principle of the optical design is shown in Fig. 2. We used the
fibre-lenslet technique described by AC.  This makes more efficient use of
the detector surface than the lenslet-only approach (e.g. Sauron; Bacon et
al 2001). While not quite as efficient as the image-slicing method (e.g. 3D;
Weitzel et al. 1996), it provides a high level of efficiency without the
technical risk that would be required to implement such a design.  The
original image-slicing technique of 3D could not be used for reasons of size
and optical performance, but a version using the {\it Advanced Image Slicer}
(AIS) concept would have been feasible for GMOS (Content 1997).  Although
IFUs using the AIS concept are under construction (Dubbeldam et al.  2000),
the technology, employing diamond-turned spherical optics, was deemed to
pose an unacceptable risk to GMOS's tight schedule.

\begin{figure}[tb]
\plotone{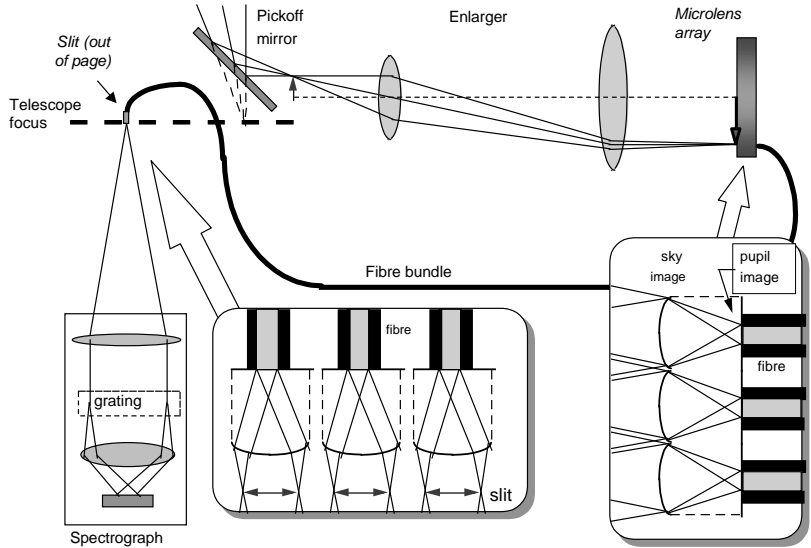}
\caption{Principle of the GMOS-IFU. The insets show details of the
fibre-lenslet coupling at the input and output of the IFU. }
\end{figure}

The design adopted folds light approaching the telescope focal plane into a
telecentric enlarger which re-images the field onto a close-packed lenslet
array. This produces an array of pupil images which are fed into optical
fibres in a fast beam to minimise the effect of focal ratio degradation
(FRD; Carrasco \& Parry 1994). This arrangement ensures that the filling
factor in the field is nearly unity since it is set by the the lenslet
array, which is close-packed, rather than by the fibres whose active area,
that of the cores, is sparsely distributed.

The fibres allow the 2-D array to be reformatted into two pseudo-slits which
are located at the original telescope focal plane. By splitting the fibres
into blocks which may be individually aligned, it is possible to mimic the
exit pupil of the telescope so that the spectrograph receives the
reformatted light as if from a pair of conventional longslits. These slit
blocks contain the fibre terminations which consist of linear lenslet arrays
which convert the beam back to the speed originally incident from the
telescope.

The confocality of the input and output focal surfaces of the IFU with that
of the telescope means that no major shift in spectrograph focus is
required when the IFU is inserted into the beam. Thus, the deployment of the
IFU is a transparent process from the point of view of the spectrograph.

In the normal two-slit mode, two sets of spectra are produced, displaced in
the dispersion direction. This allows the spectra to contain $\sim$2000
pixels but requires the use of a suitable bandpass filter to avoid overlaps
between spectra. If it is desired to increase the spectrum length and avoid
the possibility of overlaps altogether, one slit may be blocked off by the
manual insertion of an output mask.  The mapping between the field and the
slit is arranged so that this will block off one half of both object and
background field leaving contiguous sections of the field as shown in Fig.
3. This allows the observer to choose the optimum combination of the numbers
of spectral and spatial samples.

\begin{figure}[tb]
\vspace{0mm}
\plotone{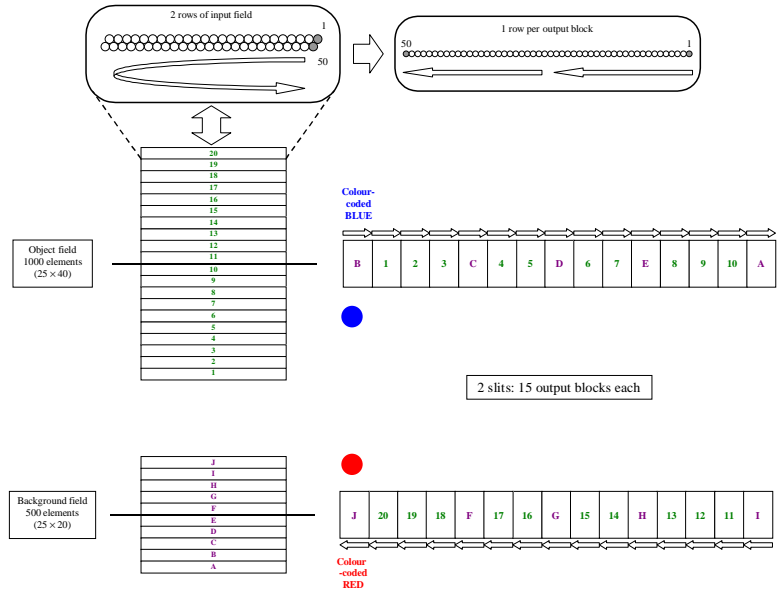}
\caption{The mapping between field and slit. The two fields are shown
schematically on the left and the two slits on the right. 
The division of each field into blocks (each
containing 2 rows at the input) is shown. The arrows indicate how the
elements in a row map onto the slit. 
Each slit maps to one contiguous half of each field,
so that the IFU can be used with one slit only to maximise spectrum length
at the expense of field area.}
\end{figure}

In order to maximise the numbers of spatial samples, the images produced by
each fibre at the pseudoslit are allowed to overlap slightly. This situation
is discussed by AC, who show that the only consequence is a slight
degradation in spatial resolution (in the direction corresponding to that of
the slit only) since the Nyquist sampling criterion is satisfied providing
that the object is sampled by at least two lenslets at the IFU input: the
sampling of the slit by the detector is only of minor importance. If the
overlap between images at the slit was minimised, the result would be a
drastic reduction in the number of spatial samples with a consequent
reduction of the field of view.  The close-packed configuration also
requires that the mapping between the input and output is such that elements
which are adjacent at the slit are also adjacent in the field, as shown in
Fig. 3, and that the element-to-element variation in throughput is small, as
described in Section 3.2.

The quality of background subtraction can be enhanced by the use of the
separate background field. This is half the size of the object field and
separated from it by 60 arcsec to allow the centre of a large extended
object to be observed while still obtaining a clean estimate of the sky
background (Fig. 4). If desired, beam-switching may be used, by which the
telescope is nodded to switch the object alternately between the object and
background field. This provides immunity to temporal variations in the sky
background, depending on the frequency of switching, and improves the
subtraction accuracy since the same physical elements sample both object and
sky. This is discussed in detail by AC.

\begin{figure}[tb]
\plotone{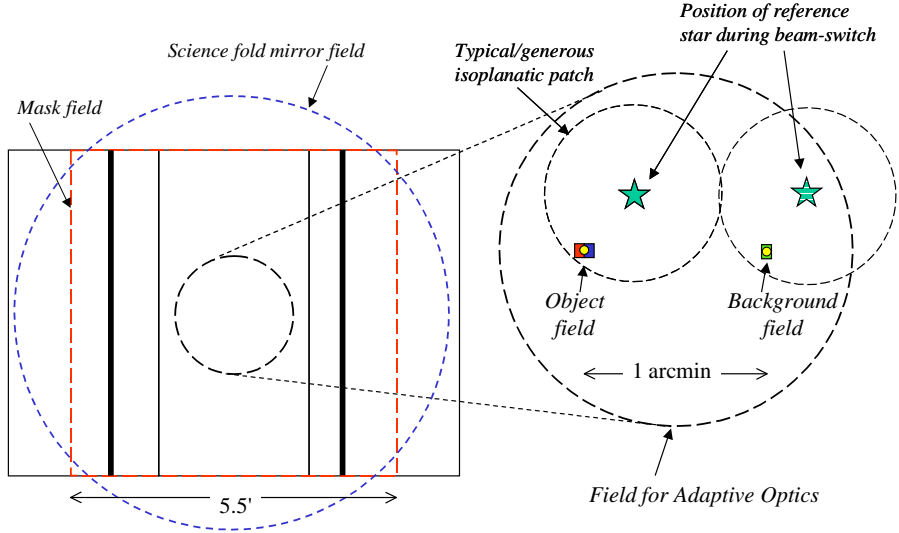}
\caption{Layout of the main GMOS and IFU fields. 
On the left the field of view
of GMOS is given by the intersection of the circular field of the science fold
mirror and the square field covered by the slit mask (both dashed). 
This is shown projected onto the detector, a mosaic of three CCDs, 
indicated by the thin solid lines. The thick lines
show the location of the IFU slits. The central portion of the GMOS
field is enlarged (right) to show the location of the two IFU fields and the
typical disposition of a reference star for both segments of a
typical beam-switch cycle, shown as stars. The field of the AO system and a
typical isoplanatic patch is also indicated.} 
\end{figure}

Fig.~5 gives an overview of the mechanical design. The fore-optics and slit
assemblies are fixed rigidly to the slit plate. The slit plate interfaces
with the mask exchange mechanism in the same way as for a regular slit mask.

\begin{figure}[tb]
\plotone{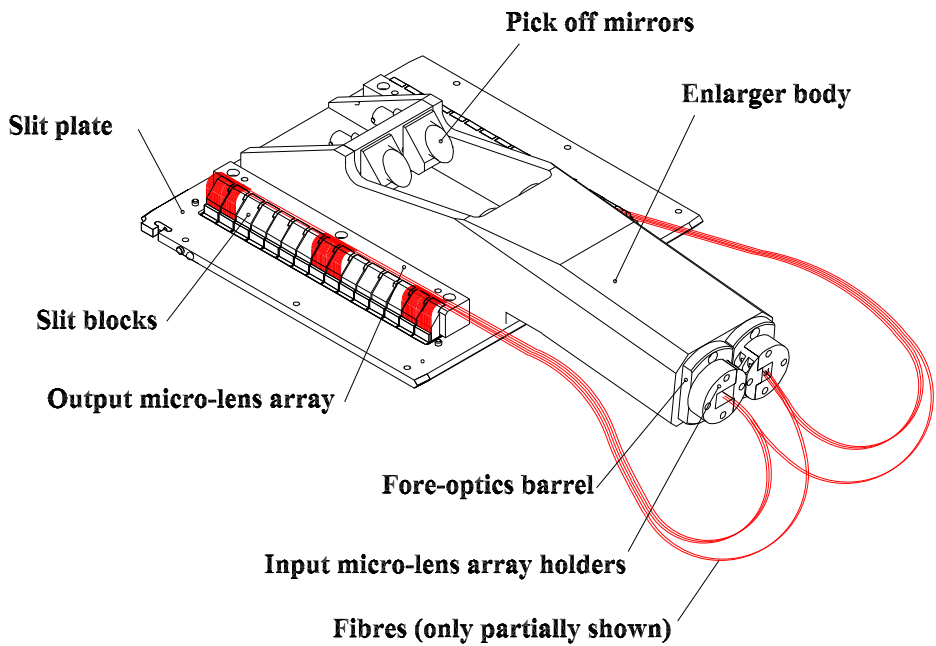}
\caption{Three-dimensional drawing of the IFU without covers. } 
\end{figure}

\section{Instrument characterisation}
\label{lab}

The performance of the IFU was determined by measurements in the laboratory
after assembly and at the telescope, in September 2001, during the
commissioning of GMOS.

\subsection{Cosmetic quality}

Out of 1500, there is only one broken element. In addition there are 4
elements with significantly reduced throughput (assessed by visual
inspection.

\subsection{Throughput and fibre-to-fibre variation}

The throughput of the IFU was measured in the laboratory using a calibrated
photodiode referenced to a wavelength of 670nm. 

The throughput of the complete IFU without regard to the spectrograph stop
(i.e. without constraining the angular extent of the output beam) was found
to be 79$\pm$5\%. The useful component of this throughput (i.e. all light
that enters the spectrograph stop) was determined to be 68$\pm$5\%. This
figure includes any slit-pointing misalignment. This compares with the
theoretical estimate made before final system optimisation of 59\%.

The throughput of the IFU after installation on the telescope was assessed
using flatfield light from the Gemini Calibration Unit (GCAL). The IFU was
deployed in GMOS and an exposure made in direct imaging mode (the grating
replaced by a plane mirror) with a particular broad-band filter in the beam.
The resulting raw data was examined and the signal integrated over the full
extent of the IFU output (the direct images of the two slits) after
subtracting the detector background.  The IFU was then removed from the beam
and another direct image taken with the same spectrograph and GCAL
configuration.  The signal was summed over those pixels which correspond to
the spatial extent of the two input fields of the IFU. A comparison of the
signal levels yields the throughput of the IFU alone over the passband
defined by the filter. The results were: 62\%, 65\%, 62\% and 58\% in the
g', r', i' and z' filter passbands respectively (SDSS system: Fukugita et
al. 1996).  The previous laboratory measurement of 68$\pm$5\% (670nm) is
consistent with the r'-band result (550-700nm).

This is the simplest and most direct method of measurement because it
is independent of variations in seeing or transparency that might
affect an observation of a standard star.

The variation in throughput from fibre to fibre was estimated from flat-field
data. Ignoring the 5 elements known to be defective, the counts recorded in
each extracted spectrum showed an RMS variation of 6\%. This is an upper
limit since the intensity of the flatfield illumination was not completely 
uniform.

\subsection{Image quality}

The same commissioning data were used to assess the quality of the images of
the individual elements at the slit, as recorded by the detector. In direct
imaging mode, these appear as a series of round images aligned in the
spatial direction. Each is clearly resolved and the images overlap at
$\sim$50\% of their peak value in the spatial direction. Fig.~6 shows a
close-up of one slit block and a cut in the spatial direction which also
shows the extent of the inter-block gap.

\begin{figure}[tb]
\vspace{0mm}
\plotone{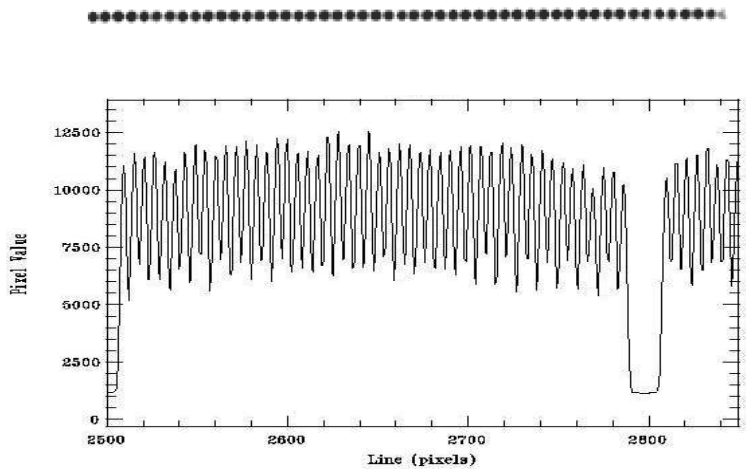}
\caption{
Top: an image of a slit block obtained with GMOS in direct-imaging mode
showing the individually resolved images of each element. 
The dispersion direction is vertical. Bottom: a cut in the spatial
direction (not to scale) also showing an inter-block gap.}
\end{figure}

\subsection{Stability}

Flexure effects were assessed by recording the positions of the images of
the IFU slit as a function of gravity vector while the telescope is placed
in different orientations. Although some flexure is seen (typically
0.1~pixel/hour), it is similar to that experienced by GMOS without the IFU
deployed (in direct-image or slit-spectroscopic modes). Thus we conclude
that any contribution from the movement of the IFU as a whole or from its
component parts is too small to be measurable.  The precision of this result
is limited by non-repeatability in the measurements. This is made up of a
combination of some intrinsic non-elasticity in the structure of GMOS and
the uncertainty in measuring the centroids of the images.

\section{Scientific characterisation}

During commissioning, a number of elliptical galaxies from the Sauron
programme (de Zeeuw et al. 2002) were observed in order to provide a
comparison dataset to allow us to verify the accuracy of the data
reduction. In addition, NGC~1068 was observed to allow a comparison with the
extensive body of long-slit spectroscopy available with e.g. STIS on the
Hubble Space Telescope (Crenshaw \& Kraemer 2000) as well as previous
integral field data, (e.g.  Thatte et al. 1997, Pecontal et al. 1997,
Garcia-Lorenzo et al. 1999).

The observations of NGC1068 consist of a set of 4 pointings offset from the
object centre by a few arcsec so as to synthesise a larger field by
mosaicing.  We used the B600 grating with the $g'$ filter to limit the
length of spectrum so as to avoid overlap between the two sets of
spectra. This configuration gave good coverage of the H$\beta$ and
[OIII]4959+5007 emission lines.

%

Fig. 7 shows a comparison of a direct image of NGC~1068 taken by GMOS with
the $g'$ filter and an image in the same passband reconstructed from the
integral field data. They are very similar, with only small differences that
are likely to be caused by the small difference between the passband of the
direct image and that used in the reduction of the spectroscopic data. This
provides a preliminary check of the accuracy of the integral field data.

\begin{figure}[tb]
\plotone{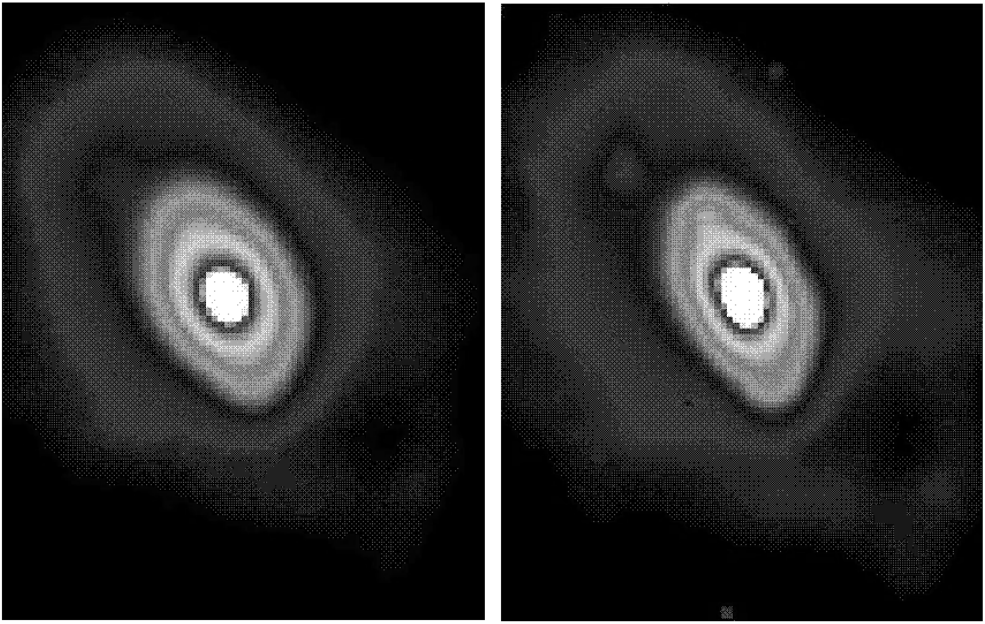}
\caption{A comparison of a direct image of the nuclear region of NGC1068 taken
with GMOS in imaging mode (left) and reconstructed from the integral field
data (right) using the same $g'$ passband filter (although the passband used
for spectrum extraction does not cover the full filter passband). The IFU
data is a mosaic of four pointings. Both
images are resampled to 0.1 arcsec/pixel. The field
shown is 10 arcsec in width. North is up and East is left.
}
\end{figure}

Fig. 8 shows how the shape and multiplicity of the emission lines changes
over the field. This confirms the initial impression of complexity obtained
from examination of the raw data.

\begin{figure}[tb]
\plotone{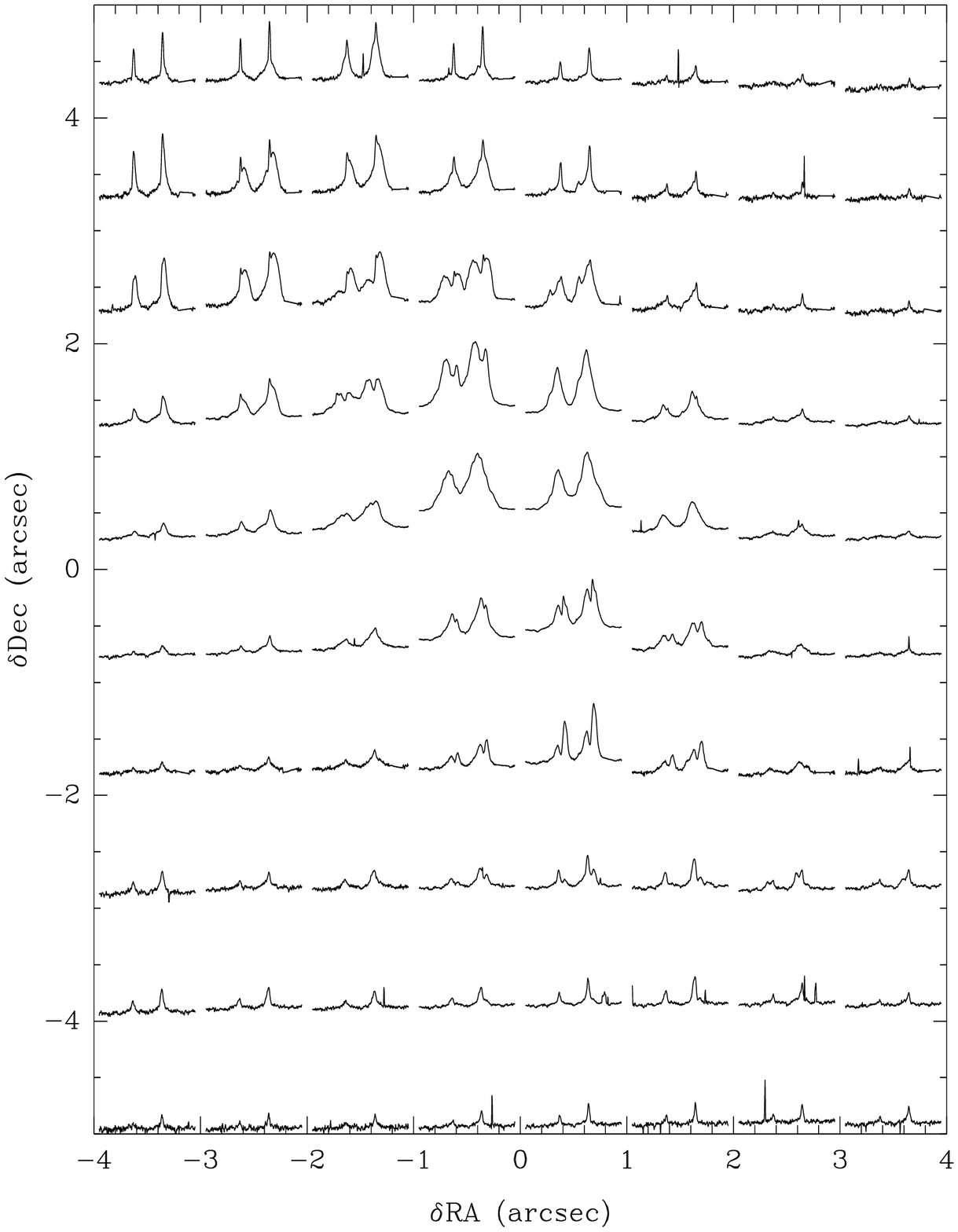}
\caption{A composite plot of the [OIII]4959+5007 emission lines 
over the field. The shape of the lines varies strongly and multiple
components are evident.}
\end{figure}

These data were reduced using a software package produced
by GEMINI for GMOS. This includes special features for the reduction of
integral field data, which is generally more complex than for slit
spectroscopy.

\section{Conclusions} 

We have described the design of the integral field unit of the Gemini
Multiobject Spectrograph installed on the northern Gemini telescope and
presented performance results obtained in the laboratory and at the
telescope. This is the first instrument for integral field spectroscopy on
an 8-10m telescope and the largest format of fibre-coupled IFU yet used at
any telescope.

Estimates of the throughput of the IFU (including losses at the spectrograph
stop) indicate reasonable agreement between theory and measurements in the
laboratory, considering that both determinations are subject to
uncertainties which are difficult to estimate. The experimentally-determined
throughput is a little higher than expected indicating the high quality of
construction and the realism of the theoretical model, which made
pessimistic assumptions where hard data was lacking. This encourages us to
believe that not only have we developed the techniques to make high-quality
devices for integral field spectroscopy but also that we have the design
tools which allows us to make a near-optimal choice of design parameters and
produce realistic estimates of the final performance.

Measurements obtained at the telescope indicate a throughput intermediate
between the laboratory and theoretical estimates, although formally
consistent with the former.  The variation in throughput between elements is
satisfactorily small and the quality of the images at the slit is actually
slightly better than predicted. This is expected to ease the task of data
reduction.

\acknowledgements

We thank the many staff in Durham who have worked on the GMOS-IFU or on its
prototypes: Roger Haynes, David Lee, Ian Lewis, Deqing Ren, Jingyun Zhang,
Ray Sharples and David Robertson.  We also thank the other members of the
GMOS team who made this work possible by supplying the superb optics,
mechanics, electronics and controls of GMOS-1, including Terry Purkins, Phil
Williams and David Lunney.  We also thank the Gemini project for supporting
this work (especially Doug Simons, Fred Gillett and Matt Mountain).

The Gemini Observatory is operated by the Association of Universities for
Research in Astronomy, Inc., under a cooperative agreement with the NSF on
behalf of the Gemini partnership: the National Science Foundation (United
States), the Particle Physics and Astronomy Research Council (United
Kingdom), the National Research Council (Canada), CONICYT (Chile), the
Australian Research Council (Australia), CNPq (Brazil) and CONICET
(Argentina).

\end{document}